\documentclass{elsart}                      

\usepackage{amsmath}
\usepackage{epsf}

\newcommand{\pythjet}{\textsc{Py\-thia/Jet\-set}}
\newcommand{\cerenkov}{\v{C}erenkov}

\newcommand{\dn}{\ensuremath{D^{\raise0.3ex\hbox{\scriptsize{\;\!0}}}}}

\newcommand{\eg}{e.g.}

\def\issue(#1,#2,#3){#1 (#3) #2} 

\def\opcit(#1){ {\em op. cit.}, #1}

\def\ARNPS(#1,#2,#3){Ann.\ Rev.\ Nucl.\ Part.\ Sci.\ \issue(#1,#2,#3)}
\def\CPC(#1,#2,#3){Comp.\ Phys.\ Comm.\ \issue(#1,#2,#3)}
\def\CIP(#1,#2,#3){Comput.\ Phys.\ \issue(#1,#2,#3)}
\def\CJP(#1,#2,#3){Chin.\ J.\ Phys.\ \issue(#1,#2,#3)}
\def\EPJC(#1,#2,#3){Eur.\ Phys.\ J.\ C\ \issue(#1,#2,#3)}
\def\IEEETNS(#1,#2,#3){IEEE Trans.\ Nucl.\ Sci.\ \issue(#1,#2,#3)}
\def\MPL(#1,#2,#3){Mod.\ Phys.\ Lett.\ \issue(#1,#2,#3)}
\def\NP(#1,#2,#3){Nucl.\ Phys.\ \issue(#1,#2,#3)}
\def\NIM(#1,#2,#3){ Nucl.\ Inst.\ and Meth.\ \issue(#1,#2,#3)}
\def\PL(#1,#2,#3){Phys.\ Lett.\ \issue(#1,#2,#3)}
\def\PRD(#1,#2,#3){Phys.\ Rev.\ D \issue(#1,#2,#3)}
\def\PRL(#1,#2,#3){Phys.\ Rev.\ Lett.\ \issue(#1,#2,#3)}
\def\SJNP(#1,#2,#3){Sov.\ J. Nucl.\ Phys.\ \issue(#1,#2,#3)}
\def\ZPC(#1,#2,#3){Zeit.\ Phys.\ C \issue(#1,#2,#3)}

\begin{document}

\begin{frontmatter}
\begin{flushright}
    \small FERMILAB-Pub-99/183-E \\ UMS/HEP/99--020~~~~~~~~~~~~
\end{flushright}
\title{	Search for Rare and Forbidden Dilepton Decays of the $D^+$, 
$D_{s}^{+}$, and $D^0$ Charmed Mesons
}
%
%
\collab {Fermilab E791 Collaboration}

\author[umis]{E.~M.~Aitala},
\author[cbpf]{S.~Amato},
\author[cbpf]{J.~C.~Anjos},
\author[fnal]{J.~A.~Appel},
\author[taun]{D.~Ashery},
\author[fnal]{S.~Banerjee},
\author[cbpf]{I.~Bediaga},
\author[umas]{G.~Blaylock},
\author[stev]{S.~B.~Bracker},
\author[stan]{P.~R.~Burchat},
\author[ilit]{R.~A.~Burnstein},
\author[fnal]{T.~Carter},
\author[cbpf]{H.~S.~Carvalho},
\author[usca]{N.~K.~Copty},
\author[umis]{L.~M.~Cremaldi},
\author[yale]{C.~Darling},
\author[fnal]{K.~Denisenko},
\author[ucin]{S.~Devmal},
\author[pueb]{A.~Fernandez},
\author[usca]{G.~F.~Fox},
\author[ucsc]{P.~Gagnon},
\author[cbpf]{C.~Gobel},
\author[umis]{K.~Gounder},
\author[fnal]{A.~M.~Halling},
\author[cine]{G.~Herrera},
\author[taun]{G.~Hurvits},
\author[fnal]{C.~James},
\author[ilit]{P.~A.~Kasper},
\author[fnal]{S.~Kwan},
\author[usca]{D.~C.~Langs},
\author[ucsc]{J.~Leslie},
\author[fnal]{B.~Lundberg},
\author[cbpf]{J.~Magnin},
\author[taun]{S.~MayTal-Beck},
\author[ucin]{B.~Meadows},
\author[cbpf]{J.~R.~T.~de~Mello~Neto},
\author[ksun]{D.~Mihalcea},
\author[tuft]{R.~H.~Milburn},
\author[cbpf]{J.~M.~de~Miranda},
\author[tuft]{A.~Napier},
\author[ksun]{A.~Nguyen},
\author[ucin,pueb]{A.~B.~d'Oliveira},
\author[ucsc]{K.~O'Shaughnessy},
\author[ilit]{K.~C.~Peng},
\author[ucin]{L.~P.~Perera},
\author[usca]{M.~V.~Purohit},
\author[umis]{B.~Quinn},
\author[uwsc]{S.~Radeztsky},
\author[umis]{A.~Rafatian},
\author[ksun]{N.~W.~Reay},
\author[umis]{J.~J.~Reidy},
\author[cbpf]{A.~C.~dos Reis},
\author[ilit]{H.~A.~Rubin},
\author[umis]{D.~A.~Sanders},
\author[ucin]{A.~K.~S.~Santha},
\author[cbpf]{A.~F.~S.~Santoro},
\author[ucin]{A.~J.~Schwartz},
\author[cine,uwsc]{M.~Sheaff},
\author[ksun]{R.~A.~Sidwell},
\author[yale]{A.~J.~Slaughter},
\author[ucin]{M.~D.~Sokoloff},
\author[cbpf]{J.~Solano},
\author[ksun]{N.~R.~Stanton},
\author[fnal]{R.~J.~Stefanski},
\author[uwsc]{K.~Stenson},  
\author[umis]{D.~J.~Summers},
\author[yale]{S.~Takach},
\author[fnal]{K.~Thorne},
\author[ksun]{A.~K.~Tripathi},
\author[uwsc]{S.~Watanabe},
\author[taun]{R.~Weiss-Babai},
\author[prin]{J.~Wiener},
\author[ksun]{N.~Witchey},
\author[yale]{E.~Wolin},
\author[ksun]{S.~M.~Yang},
\author[umis]{D.~Yi},
\author[ksun]{S.~Yoshida},
\author[stan]{R.~Zaliznyak}, and
\author[ksun]{C.~Zhang}

\address[cbpf]{Centro Brasileiro de Pesquisas F{\'\i}sicas, Rio de 
Janeiro, Brazil}
\address[ucsc]{University of California, Santa Cruz, California 
95064, USA}
\address[ucin]{University of Cincinnati, Cincinnati, Ohio 45221, USA}
\address[cine]{CINVESTAV, 07000 Mexico City, DF Mexico}
\address[fnal]{Fermilab, Batavia, Illinois 60510, USA}
\address[ilit]{Illinois Institute of Technology, Chicago, Illinois 
60616, USA}
\address[ksun]{Kansas State University, Manhattan, Kansas 66506, USA}
\address[umas]{University of Massachusetts, Amherst, Massachusetts 
01003, USA}
\address[umis]{University of Mississippi--Oxford, University, 
Mississippi 38677, USA}
\address[prin]{Princeton University, Princeton, New Jersey 08544, USA}
\address[pueb]{Universidad Autonoma de Puebla, Mexico}
\address[usca]{University of South Carolina, Columbia, South Carolina 
29208, USA}
\address[stan]{Stanford University, Stanford, California 94305, USA}
\address[taun]{Tel Aviv University, Tel Aviv 69978, Israel}
\address[stev]{Box 1290, Enderby, British Columbia V0E 1V0, Canada}
\address[tuft]{Tufts University, Medford, Massachusetts 02155, USA}
\address[uwsc]{University of Wisconsin, Madison, Wisconsin 53706, USA}
\address[yale]{Yale University, New Haven, Connecticut 06511, USA}

\begin{abstract}
We report the results of a search for flavor-changing neutral current, 
lepton-flavor violating, and lepton-number violating decays of
$D^+$, $D_{s}^{+}$, and $D^0$ mesons (and their antiparticles) into 
modes containing muons and electrons. Using data from Fermilab charm 
hadroproduction experiment E791, we examine the $\pi \ell \ell$ and 
$K\ell \ell$ decay modes of $D^+$ and $D_{s}^{+}$ and the 
$\ell^+ \ell^-$ decay modes of $D^0$. No evidence for any of these 
decays is found. Therefore, we present branching-fraction upper limits 
at 90$\%$ confidence level for the 24 decay modes examined. Eight of 
these modes have no previously reported limits, and fourteen are 
reported with significant improvements over previously published 
results.
\end{abstract}
\end{frontmatter}
The SU(2)$\times $U(1) Standard Model of electroweak interactions
qualitatively accounts for the known decays of heavy quarks and can
often quantitatively predict the decay rates. However, this model
is incomplete in that it does not account for the number of quark and 
lepton families observed, nor their hierarchy of mass scales. Also 
unknown is the mechanism responsible for breaking the underlying gauge 
symmetry. One way to search for physics beyond the Standard Model is to 
search for decays that are forbidden or else are predicted to occur at 
a negligible level. Observing such decays would constitute evidence for 
new physics, and measuring their branching fractions would provide 
insight into how to modify our theoretical understanding, \eg, by 
introducing new particles or new gauge couplings.

In this letter we present the results of a search for 24 decay modes of 
the neutral and charged $D$ mesons (which contain the heavy charm 
quark). These decay modes\footnote{Charge-conjugate modes are 
included implicitly throughout this paper.} fall into three categories:
\begin{enumerate}
\item FCNC -- flavor-changing neutral current decays 
($D^0\rightarrow\ell^+\ell^-$ and
$D^+_{(d,s)}\rightarrow h^+\ell^+\ell^-$, in which $h$ is
$\pi$ or $K$);
\item LFV -- lepton-flavor violating decays 
($D^{0}\rightarrow \mu^{\pm }e^{\mp }$,
$D^+_{(d,s)}\rightarrow h^+\mu^{\pm }e^{\mp }$, and 
$D^+_{(d,s)}\rightarrow h^-\mu^+e^+$,
in which the leptons belong to different generations);
\item LNV -- lepton-number violating decays
($D^+_{(d,s)}\rightarrow h^-\ell^+\ell^+$,
in which the leptons belong to the same generation
but have the same sign charge).
\end{enumerate}
Decay modes belonging to (1) occur within the Standard Model
via higher-order diagrams, but the estimated branching fractions are
$10^{-8}$ to $10^{-6}$ \cite{SCHWARTZ93}. Such small rates are 
below the sensitivity of current experiments. However, if 
additional particles such as supersymmetric squarks or charginos exist, 
they could contribute additional amplitudes that would make these modes 
observable. Decay modes belonging to (2) and (3) do not conserve lepton 
number and thus are forbidden within the Standard Model. However, lepton 
number conservation is not required by Lorentz invariance or gauge 
invariance, and a number of theoretical extensions to the Standard 
Model predict lepton-number violation \cite{Pakvasa}. Many experiments 
have searched for lepton-number violation in $K$ decays, and for 
lepton-number violation and flavor-changing 
neutral currents in $D$ and $B$  
decays. The limits we present here for rare and forbidden dilepton 
decays of the $D$ mesons are typically more stringent than those 
obtained from previous searches \cite{PDG}, or else are the first
reported.

The data are from Fermilab experiment E791 \cite{e791spect}, which 
recorded $2 \times 10^{10}$ events with a loose transverse energy 
trigger. These events were produced by a 500 GeV/$c$~ $\pi ^{-}$ beam 
interacting in a target consisting of five thin foils that had 15 mm 
center-to-center separation along the beamline. The most upstream foil 
was 0.5 mm thick platinum. It was followed by 
four foils consisting of 1.6 mm thick diamond.
Momentum analysis was provided by two dipole magnets that bent 
particles in the horizontal ($x$-$z$) plane. Position information for 
track and vertex reconstruction was provided by 23 silicon microstrip 
detectors (6 upstream and 17 downstream of the target) along with 10 
planes of proportional wire chambers (8 upstream and 2 downstream of 
the target), and 35 drift chamber planes. The experiment also included 
electromagnetic and hadronic calorimeters, a muon detector, and two 
multi-cell \cerenkov{} counters that provided $\pi /K$ separation in 
the momentum range $6-60$~GeV/$c$ \cite{Bartlett}. The kaon 
identification criteria varied by search decay mode.  We typically 
required that the momentum-dependent light yield in the \cerenkov{} 
counters be consistent with that of a kaon track measured in the 
spectrometer.

Electrons were identified by an electromagnetic calorimeter~\cite{SLIC} 
that consisted of lead sheets and liquid scintillator located 19~m 
downstream of the target. Electron identification was based on energy 
deposition  and transverse shower shape in the calorimeter.  The 
electron identification efficiency varied from 62$\%$ for momenta 
below 9 GeV/$c$ to 45$\%$ for momenta above 20 GeV/$c$. The decrease 
in efficiency with increasing momentum reflects the fact that higher 
momentum electrons populate a more congested region of the 
spectrometer. The pion misidentification rate was approximately 
0.8$\%$, independent of pion momentum.

Muon identification was obtained from two planes of scintillation 
counters. The plane that measured vertical coordinates ($y$) 
consisted of 16 scintillation counters, each 3 meters long and 14~cm 
wide. The plane that measured horizontal coordinates ($x$) consisted 
of 14 counters, each 3 meters long and covering a full width 
of 5.5 meters in the $x$-direction. The counters were located behind 
shielding with a thickness equivalent to 2.5 meters (15 interaction 
lengths) of iron. Candidate muon tracks projected into the muon system 
were required to pass a series of muon quality criteria that were 
optimized with 
$D^+\rightarrow \overline{K}^{*0} \mu^{+}\nu _{\!_{\mu}}$ decays from 
our data \cite{Chong}. Timing information from the $y$-coordinate 
counters was used to improve the position resolution in the 
$x$-direction. The efficiencies of the muon counters were 
measured in special runs using muons originating from the primary beam 
dump, and were found to be $(99\pm 1)\%$ for the $y$-coordinate counters 
and $(69\pm 3)\%$ for the $x$-coordinate counters. The probability for 
misidentifying a pion as a muon decreased as momentum increased, from 
about 6$\%$ at 8 GeV/$c$ to $(1.3 \pm 0.1)\%$ for momenta greater than 
20 GeV/$c$.

After reconstruction, events with evidence of well-separated 
production (primary) and decay (secondary) vertices were retained for 
further analysis. To separate charm candidates from background, we 
required the following: that secondary vertices be well-separated from 
the primary vertex and located well outside the target foils and other 
solid material; that the momentum vector of the candidate charm meson 
point back to the primary vertex; and that the decay track candidates 
pass approximately 10 times closer to the secondary vertex than to the 
primary vertex. A secondary vertex had to be separated from the primary 
vertex by greater than $20\,\sigma_{_{\!L}}$ for $D^+$ decays and 
greater than $12\,\sigma_{_{\!L}}$ for $D^0$ and $D_{s}^{+}$ decays, 
where $\sigma_{_{\!L}}$ is the calculated resolution of the measured 
longitudinal separation. In addition, the secondary vertex had to be 
separated from the closest material in the target foils by greater than 
$5\,\sigma_{_{\!L}}^{\prime }$, where $\sigma_{_{\!L}}^{\prime }$ is 
the uncertainty in this separation. The sum of the vector momenta of 
the tracks from the secondary vertex was required to pass within 
$40~\mu$m of the primary vertex in the plane perpendicular to the beam. 
Finally, the net momentum of the charm candidate transverse to 
the line connecting the production and decay vertices had to be 
less than 300 MeV/$c$ for $D^0$ candidates,
less than 250 MeV/$c$ for $D_{s}^{+}$ candidates, and
less than 200 MeV/$c$ for $D^+$ candidates. These selection criteria 
and, where possible, the kaon identification requirements, were the same 
for the search mode and for its normalization signal.

For this study we used a ``blind'' analysis technique. Before our 
selection criteria were finalized, all events having masses within a 
mass window $\Delta M_S$ around the mass of $D^{+}$, 
$D_{s}^{+}$, or $D^{0}$ were ``masked'' so that the presence or 
absence of any potential signal candidates would not bias our choice of 
selection criteria. All criteria were then chosen by studying 
signal events generated by a Monte Carlo simulation program (see 
below) and background events from real data. Events within the signal 
windows were unmasked only after this optimization. Background events 
were chosen from a mass window $\Delta M_B$ above and below the signal 
window $\Delta M_S$. The criteria were chosen to maximize the ratio 
$N_S/\sqrt{N_B}$, where $N_S$ and $N_B$ are the numbers of signal and 
background events, respectively. We used asymmetric windows for the 
decay modes containing electrons to allow for the bremsstrahlung 
low-energy tail. The signal windows are: 
\begin{equation}
     \begin{array}{l}
         1.84<M(D^{+})<1.90~{\rm GeV}/c^{\,2}~{\rm for}~ 
	 D^{+}\rightarrow h\mu \mu ,\\ 
         1.78<M(D^{+})<1.90 ~{\rm GeV}/c^{\,2}~{\rm for}~ 
	 D^{+}\rightarrow hee ~{\rm and}~  h\mu e,\\
	 1.95<M(D_{s}^{+})<1.99 ~{\rm GeV}/c^{\,2}~{\rm for}~ 
	 D_{s}^{+}\rightarrow h\mu \mu,\\
	 1.91<M(D_{s}^{+})<1.99 ~{\rm GeV}/c^{\,2}~{\rm for}~ 
	 D_{s}^{+}\rightarrow hee~{\rm and}~ h\mu e,\\
	 1.83<M(D^{0})<1.90 ~{\rm GeV}/c^{\,2}~{\rm for}~ 
	 D^{0}\rightarrow \mu \mu ,\\
	 1.76<M(D^{0})<1.90 ~{\rm GeV}/c^{\,2}~{\rm for}~ 
	 D^{0}\rightarrow ee~{\rm and}~ \mu e.
    \end{array}
\end{equation}

\begin{figure}[ht]
\vskip -2.0 cm
\centerline{\epsfxsize 5.0 truein \epsfbox{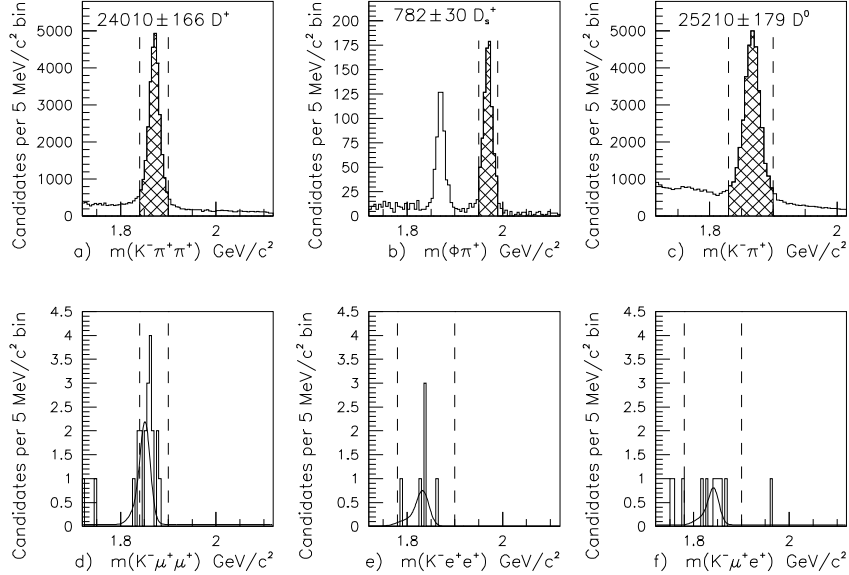}}
\vskip -.2 cm
\caption[]{
\small 
Top row: typical charm signals in normalization modes used for the a) 
$D^+$, b) $D_{s}^{+}$, and c) $D^0$ decay modes. The signal region is 
shaded. 
Bottom row: invariant mass plots of $D^{+}$ candidate decays to 
d) $K^-\mu ^+\mu ^+$, e) $K^-e^+e^+$, and f) $K^-\mu ^+e^+$, showing 
reflections primarily from misidentified $D^+\rightarrow 
K^-\pi^+\pi^+$ decays. These modes are not used to set upper limits 
but are instead used to estimate misidentification rates following 
the method described in the text. The solid curves are normalized 
Monte Carlo fits. The dashed lines show the signal window.
}
\label{Norm}
\end{figure}

We normalize the sensitivity of our search to topologically similar 
Cabibbo-favored decays. For the $D^{+}$ decays we use 
$D^+\rightarrow K^-\pi^+\pi^+$; for $D_{s}^{+}$ decays we use 
$D_{s}^{+}\rightarrow \phi \pi^+$; and for $D^{0}$ decays we 
use $D^0\rightarrow K^-\pi^+$. The widths of our normalization modes 
were 10.5 MeV/$c^{\,2}$ for $D^{+}$, 9.5 MeV/$c^{\,2}$ for $D_{s}^{+}$, 
and 12 MeV/$c^{\,2}$ for $D^{0}$. The events within the 
$\sim 5\,\sigma $ window are shown in Figs. \ref{Norm}a--c. 
The upper limit for each branching fraction $B_{X}$ is calculated using 
the following formula:
\begin{equation}
B_{X}=\frac{N_{X}}{N_{\mathrm{Norm}}}
\frac{\varepsilon _{\mathrm{Norm}}}{\varepsilon _{X}}
\cdot B_{\mathrm{Norm}} 
\label{BReqn}
\end{equation}
where $N_{X}$ is the 90$\%$ CL upper limit on the number of decays 
for the rare or forbidden decay mode $X$, and $\varepsilon_{X}$ is that 
mode's detection efficiency. $N_{\mathrm{Norm}}$ is the fitted number 
of normalization mode decays; $\varepsilon_{\mathrm{Norm}}$ 
is the normalization mode detection efficiency; and 
$B_{\mathrm{Norm}}$ is the normalization mode branching fraction 
obtained from the Particle Data Group \cite{PDG}. 

The ratio of detection efficiencies is 
\begin{equation}
\frac{\varepsilon _{\mathrm{Norm}}}{\varepsilon _{X}}=
\frac{N_{\mathrm{Norm}}^{\mathrm{MC}}}{N_{X}^{\mathrm{MC}}}
\label{Effyeqn}
\end{equation}
where $N_{\mathrm{Norm}}^{\mathrm{MC}}$ and $N_{X}^{\mathrm{MC}}$ are 
the fractions of Monte Carlo events that are reconstructed and pass 
the final selection criteria, for the normalization and decay modes 
respectively. The simulations use \pythjet~\cite{MC} as the physics 
generator and model the effects of resolution, geometry, magnetic 
fields, multiple scattering, interactions in the detector material, 
detector efficiencies, and the analysis selection criteria. The 
efficiencies for the normalization modes varied from approximately 
$0.5\%$ to $2\%$ depending on the mode, and the efficiencies for the 
search modes varied from approximately $0.1\%$ to $2\%$. 

Monte Carlo studies show that the experiment's acceptances are nearly 
uniform across the Dalitz plots, except that the dilepton 
identification efficiencies typically drop to near zero at the 
dilepton mass threshold. While the loss in efficiency varies channel by 
channel, the efficiency typically reaches its full value at masses 
only a few hundred MeV/$c^{\,2}$ above the dilepton mass threshold. We 
use a constant weak-decay matrix element when calculating the overall 
detection efficiencies. Two exceptions to the use of the Monte Carlo 
simulations in determining relative efficiencies are made: those for 
\cerenkov{} identification when the number of kaons in the signal and 
normalization modes are different, and those for the muon 
identification. These efficiencies are determined from data.

\begin{figure}[hb]
\vskip -2.5 cm
\centerline{\epsfxsize 5.0 truein \epsfbox{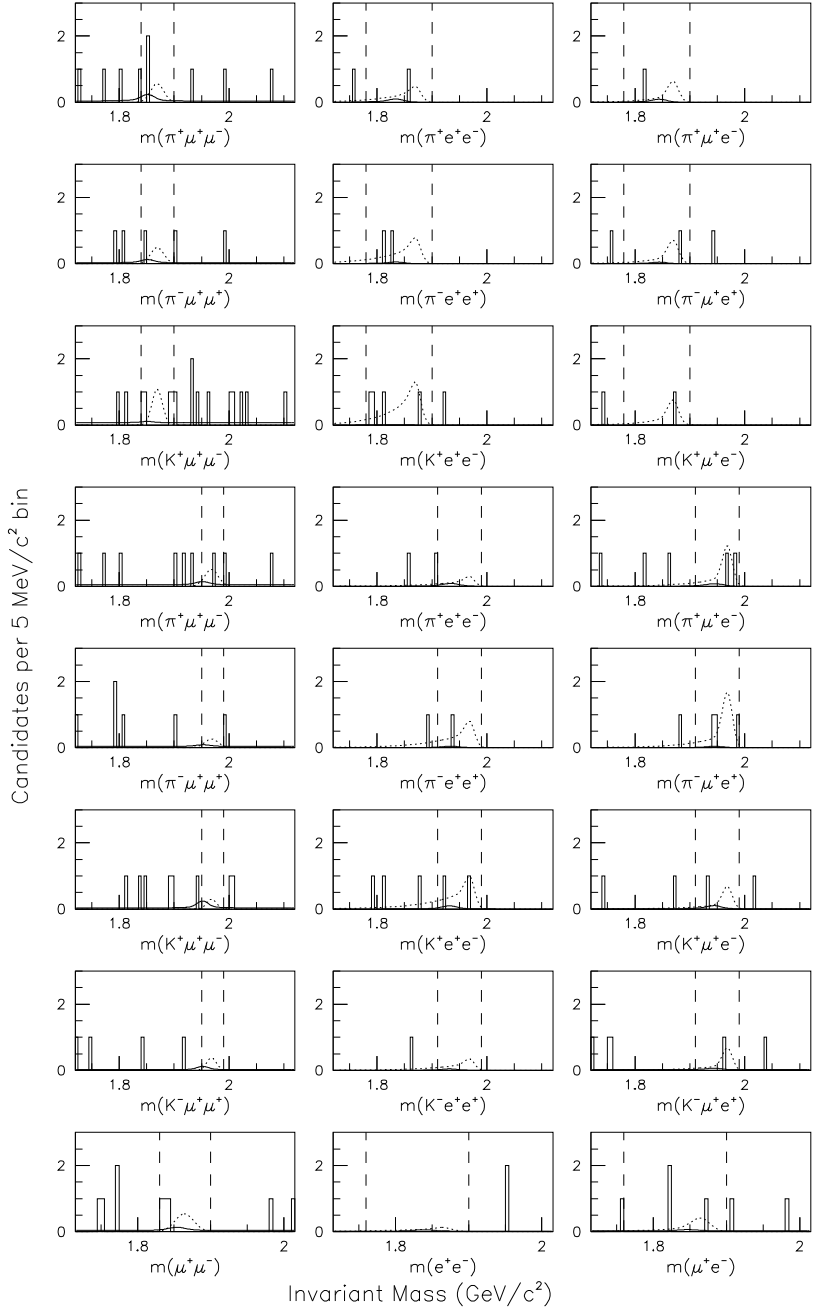}}
\vskip -.2 cm
\caption[]{
\small Final event samples for the $D^+$ (rows 1--3), 
$D_{s}^{+}$ (rows 4--7), and $D^0$ (row 8) decays. The solid curves 
represent estimated background; the dotted curves represent signal 
shape for a number of events equal to the 90$\%$ CL upper limit. 
The dashed vertical lines are $\Delta M_S$ boundaries.}
\label{Data}
\end{figure}
The 90$\%$ CL upper limits $N_{X}$ are calculated using the method of 
Feldman and Cousins \cite{Cousins} to account for background, and then 
corrected for systematic errors by the method of Cousins and Highland 
\cite{COUSINSHI}. In these methods, the numbers of signal events are 
determined by simple counting, not by a fit. All results are listed in 
Table \ref{Results} and shown in Figs. \ref{Data} and \ref{BR}. The 
kinematic criteria and removal of reflections (see below) are different 
for the $D^{+}$, $D_{s}^{+}$, and $D^0$. Thus, the $D^+$ and 
$D_{s}^{+}$ rows in Fig. \ref{Data} with the same decay particles are 
different, and the seventh row of Fig. \ref{Data} is different from the 
bottom row of Fig. \ref{Norm}.

The upper limits are determined by both the number of candidate events 
and the expected number of background events within the signal region.  
Background sources that are not removed by the selection 
criteria discussed earlier include decays in which hadrons (from 
real, fully-hadronic decay vertices) are misidentified as leptons. 
In the case where kaons are misidentified as leptons, candidates 
have effective masses which lie outside the signal windows. Most of 
these originate from Cabibbo-favored modes 
$D^+\rightarrow K^-\pi^+\pi^+$, $D_{s}^{+}\rightarrow K^-K^+\pi^+$, 
and $D^0\rightarrow K^-\pi^+$ (and charge conjugates). These 
Cabibbo-favored reflections are explicitly removed prior to the 
selection-criteria optimization. There remain two sources of background 
in our data: hadronic decays with pions misidentified as leptons 
($N_{\mathrm{MisID}}$) and ``combinatoric'' background 
($N_{\mathrm{Cmb}}$) arising primarily from false vertices and 
partially reconstructed charm decays. After selection criteria 
were applied and the signal windows opened, the number of events 
within the window is $N_{\mathrm{Obs}} = N_{\mathrm{Sig}}+ 
N_{\mathrm{MisID}} + N_{\mathrm{Cmb}}$. 

The background $N_{\mathrm{MisID}}$ arises mainly 
from singly-Cabibbo-suppressed (SCS) modes. These misidentified 
leptons can come from hadronic showers 
reaching the muon counter, decays-in-flight, and random overlaps of 
tracks from otherwise separate decays (``accidental'' sources). We do 
not attempt to establish a limit for $D^+\rightarrow K^-\ell^+\ell^+$ 
modes, as they have relatively large feedthrough signals from copious 
Cabibbo-favored $K^-\pi^+\pi^+$ decays. Instead, we use the observed 
signals in $K^-\ell^+\ell^+$ channels to measure three dilepton 
misidentification rates under the assumption that the observed signals 
(shown in Figs.~\ref{Norm}d--f) arise entirely from lepton 
misidentification. The curve shapes were determined from Monte Carlo 
simulations. The following misidentification rates were obtained: 
$r_{\mu\mu}= (7.3 \pm 2.0)\times 10^{-4}$, 
$r_{\mu e}= (2.9 \pm 1.3 )\times 10^{-4}$, and 
$r_{e e}= (3.4 \pm 1.4)\times 10^{-4}$. 
Using these rates we estimate the numbers of misidentified candidates, 
$N_{\mathrm{MisID}}^{h\ell\ell}$ (for $D^{+}$ and $D_{s}^{+}$) and 
$N_{\mathrm{MisID}}^{\ell\ell}$ (for $D^{0}$), in the signal windows 
as follows:
\begin{equation}
N_{\mathrm{MisID}}^{h\ell\ell} = r_{\ell\ell} 
\cdot N_{\mathrm{SCS}}^{h\pi\pi}
~~~{\rm and}~~~ N_{\mathrm{MisID}}^{\ell\ell} = r_{\ell\ell} 
\cdot N_{\mathrm{SCS}}^{\pi\pi}~,
\label{Nmidid}
\end{equation}
where $N_{\mathrm{SCS}}^{h\pi\pi}$ and $N_{\mathrm{SCS}}^{\pi\pi}$ 
are the numbers of SCS hadronic decay candidates within the signal 
windows. For modes in which two possible pion combinations can 
contribute, \eg, $D^+\rightarrow h^{+}\mu ^{\pm}\mu ^{\mp}$, we use 
twice the above rate. These misidentification backgrounds were 
typically small or negligible.

To estimate the combinatoric background $N_{\mathrm{Cmb}}$ within a 
signal window $\Delta M_S$, we count events having masses within an 
adjacent background mass window $\Delta M_B$, and scale this number 
($N_{\Delta M_B}$) by the relative sizes of these windows:
\begin{equation}
N_{\mathrm{Cmb}} = \frac{\Delta M_S}{\Delta M_B} \cdot N_{\Delta M_B}. 
\label{Nbak}
\end{equation}
To be conservative in calculating our 90$\%$ confidence level upper 
limits, we take combinatoric backgrounds to be zero when no 
events are located above the mass windows. In Table \ref{Results} we 
present the numbers of combinatoric background, misidentification 
background, and observed events for all 24 modes.

The sources of systematic errors in this analysis include: 
statistical errors from the fit to the normalization sample 
$N_{\mathrm{Norm}}$; statistical errors on the numbers of Monte Carlo 
generated events for both $N_{\mathrm{Norm}}^{\mathrm{MC}}$ and 
$N_{X}^{\mathrm{MC}}$; uncertainties in the calculation of 
misidentification background; and uncertainties in the relative 
efficiency for each mode, including lepton and kaon tagging 
efficiencies. These tagging efficiency uncertainties include: 1) the 
muon counter efficiencies from both Monte Carlo simulation and 
hardware performance; 2) kaon \cerenkov{} identification 
efficiency due to differences in kinematics and modeling between 
data and Monte Carlo simulated events; and 3) the fraction of 
signal events (based on simulations) that would remain outside the 
signal window due to bremsstrahlung tails. 
The larger systematic errors for the $D_{s}^{+}$ modes, 
compared to the $D^{+}$ and $D^{0}$ modes, are due to the uncertainty 
in the branching fraction for the $D_{s}^{+}$ normalization mode.
The sums, taken in quadrature, of these systematic errors are listed 
in Table \ref{Results}.
\begin{table}[ht]
\caption[]{
E791 90$\%$ confidence level (CL) branching fractions (BF) compared 
to PDG98 limits. The background and candidate events 
correspond to the signal region only.}
\label{Results}
\vskip 5pt
\tabcolsep=4.0pt
\begin{tabular}{lcccccll} \hline
&(Est.&BG)&Cand.&Syst.&90$\%$ CL&E791&PDG98 \cite{PDG} \\
Mode&$N_{\mathrm{Cmb}}$&$N_{\mathrm{MisID}}$&Obs.&Err.&Num.
&$BF$ Limit&$BF$ Limit \\
\hline
 $D^{+}\rightarrow \pi ^{+}\mu ^{+}\mu ^{-}$&1.20&1.47&2&10$\%$&3.35
 &$1.5\times 10^{-5}$&$1.8\times 10^{-5}$\\
 $D^{+}\rightarrow \pi ^{+}e^{+}e^{-}$&0.00&0.90&1&12$\%$&3.53
 &$5.2\times 10^{-5}$&$6.6\times 10^{-5}$\\
 $D^{+}\rightarrow \pi ^{+}\mu ^{\pm }e^{\mp }$&0.00&0.78&1&11$\%$&3.64
 &$3.4\times 10^{-5}$&$1.2\times 10^{-4}$\\
 $D^{+}\rightarrow \pi ^{-}\mu ^{+}\mu ^{+}$&0.80&0.73&1&9$\%$&2.92
 &$1.7\times 10^{-5}$&$8.7\times 10^{-5}$\\
 $D^{+}\rightarrow \pi ^{-}e^{+}e^{+}$&0.00&0.45&2&12$\%$&5.60
 &$9.6\times 10^{-5}$&$1.1\times 10^{-4}$\\
 $D^{+}\rightarrow \pi ^{-}\mu ^{+}e^{+}$&0.00&0.39&1&11$\%$&4.05
 &$5.0\times 10^{-5}$&$1.1\times 10^{-4}$\\
 $D^{+}\rightarrow K^{+}\mu ^{+}\mu ^{-}$&2.20&0.20&3&8$\%$&5.07
 &$4.4\times 10^{-5}$&$9.7\times 10^{-5}$\\
 $D^{+}\rightarrow K^{+}e^{+}e^{-}$&0.00&0.09&4&11$\%$&8.72
 &$2.0\times 10^{-4}$&$2.0\times 10^{-4}$\\
 $D^{+}\rightarrow K^{+}\mu ^{\pm }e^{\mp }$&0.00&0.08&1&9$\%$&4.34
 &$6.8\times 10^{-5}$&$1.3\times 10^{-4}$\\
\hline
 $D_{s}^{+}\rightarrow K^{+}\mu ^{+}\mu ^{-}$&0.67&1.33&0&27$\%$&1.32
 &$1.4\times 10^{-4}$&$5.9\times 10^{-4}$\\
 $D_{s}^{+}\rightarrow K^{+}e^{+}e^{-}$&0.00&0.85&2&29$\%$&5.77
 &$1.6\times 10^{-3}$&\\
 $D_{s}^{+}\rightarrow K^{+}\mu ^{\pm }e^{\mp }$&0.40&0.70&1&27$\%$&3.57
 &$6.3\times 10^{-4}$&\\
 $D_{s}^{+}\rightarrow K^{-}\mu ^{+}\mu ^{+}$&0.40&0.64&0&26$\%$&1.68
 &$1.8\times 10^{-4}$&$5.9\times 10^{-4}$\\
 $D_{s}^{+}\rightarrow K^{-}e^{+}e^{+}$&0.00&0.39&0&28$\%$&2.22
 &$6.3\times 10^{-4}$&\\
 $D_{s}^{+}\rightarrow K^{-}\mu ^{+}e^{+}$&0.80&0.35&1&27$\%$&3.53
 &$6.8\times 10^{-4}$&\\
 $D_{s}^{+}\rightarrow \pi ^{+}\mu ^{+}\mu ^{-}$&0.93&0.72&1&27$\%$&3.02
 &$1.4\times 10^{-4}$&$4.3\times 10^{-4}$\\
 $D_{s}^{+}\rightarrow \pi ^{+}e^{+}e^{-}$&0.00&0.83&0&29$\%$&1.85
 &$2.7\times 10^{-4}$&\\
 $D_{s}^{+}\rightarrow \pi ^{+}\mu ^{\pm }e^{\mp }$&0.00&0.72&2&30$\%$
 &6.01&$6.1\times 10^{-4}$&\\
 $D_{s}^{+}\rightarrow \pi ^{-}\mu ^{+}\mu ^{+}$&0.80&0.36&0&27$\%$&1.60
 &$8.2\times 10^{-5}$&$4.3\times 10^{-4}$\\
 $D_{s}^{+}\rightarrow \pi ^{-}e^{+}e^{+}$&0.00&0.42&1&29$\%$&4.44
 &$6.9\times 10^{-4}$&\\
 $D_{s}^{+}\rightarrow \pi ^{-}\mu ^{+}e^{+}$&0.00&0.36&3&28$\%$&8.21
 &$7.3\times 10^{-4}$&\\
\hline
 $D^{0}\rightarrow \mu ^{+}\mu ^{-}$&1.83&0.63&2&6$\%$&3.51
 &$5.2\times 10^{-6}$&$4.1\times 10^{-6}$\\
 $D^{0}\rightarrow e^{+}e^{-}$&1.75&0.29&0&9$\%$&1.26
 &$6.2\times 10^{-6}$&$1.3\times 10^{-5}$\\
 $D^{0}\rightarrow \mu ^{\pm }e^{\mp }$&2.63&0.25&2&7$\%$&3.09
 &$8.1\times 10^{-6}$&$1.9\times 10^{-5}$\\
\hline
\end{tabular}
\end{table}
\vfill
\eject

In summary, we use a ``blind'' analysis of data from Fermilab 
experiment E791 to obtain upper limits on the dilepton branching 
fractions for flavor-changing neutral current, lepton-number violating, 
and lepton-family violating decays of $D^+$, $D_{s}^{+}$, and $D^0$ 
mesons. No evidence for any of these decays is found. Therefore, we 
present upper limits on the branching fractions at the 90$\%$ 
confidence level. These limits represent significant improvements over 
previously published results. Eight new $D_{s}^{+}$ search modes are 
reported. A comparison of our 90$\%$ C.L. upper limits with previously 
published results \cite{PDG} is shown in Fig. \ref{BR}.
\begin{figure}[hb]
\centerline{\epsfxsize 5.0 truein \epsfbox{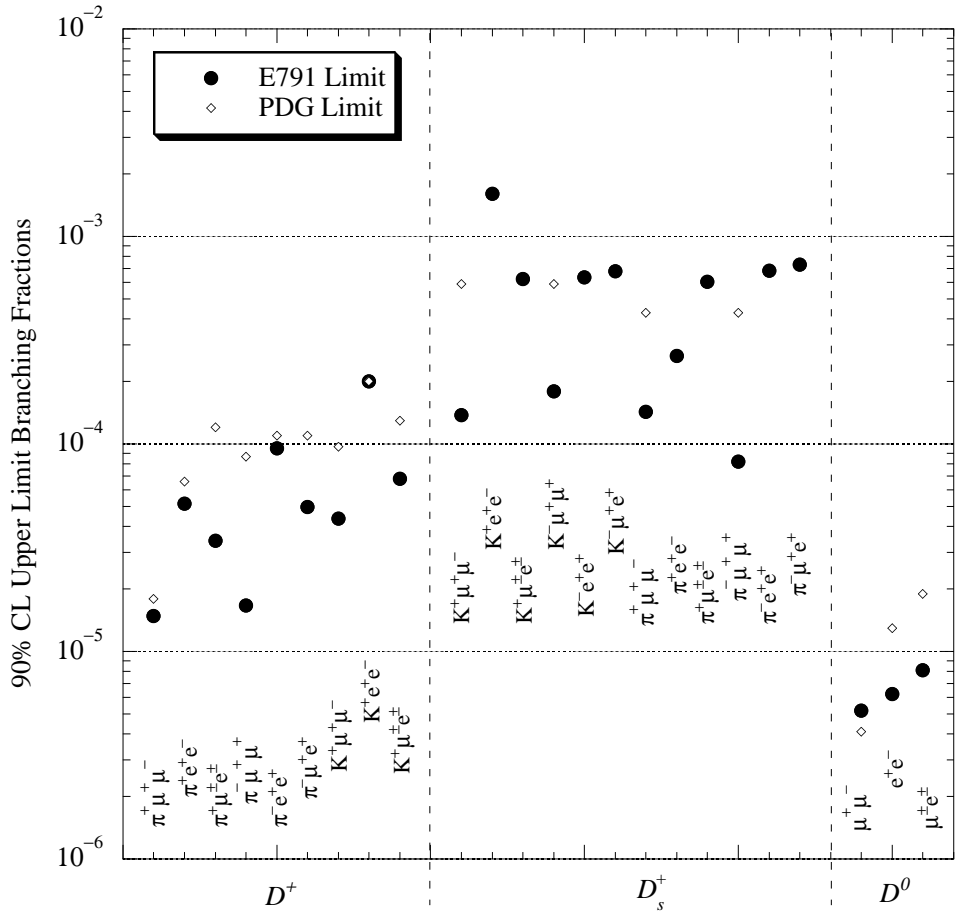}}
\caption[]{
\small Comparison of the 90$\%$ CL upper-limit branching fractions 
from E791 data (dark circles) with existing limits (open diamonds) from 
the 1998 PDG \cite{PDG}.}
\label{BR}
\end{figure}

We gratefully acknowledge the assistance of the staffs of Fermilab and
of all the participating institutions. This research was supported by
the Brazilian Conselho Nacional de Desenvolvimento Cient\'\i fico e
Tecnol\'ogico, CONACyT (Mexico), the Israeli Academy of Sciences and
Humanities, the U.S. Department of Energy, the U.S.-Israel Binational
Science Foundation, and the U.S. National Science Foundation. Fermilab
is operated by the Universities Research Association, Inc., under
contract with the U.S. Department of Energy.
\vfill
\eject

\bibliographystyle{unsrt}

\end{document}